\documentclass[aps,twocolumn,pra,showpacs,floatfix]{revtex4}
\usepackage{epsfig}
\usepackage{dcolumn}

\begin{document}

\title{\bf Optical clock sensitive to variation of the fine structure
constant based on the Ho$^{14+}$ ion}

\author{V. A. Dzuba and V. V. Flambaum}
\affiliation{School of Physics, University of New South Wales, 
Sydney 2052, Australia}

\author{Hidetoshi Katori} 
\affiliation{
Quantum Metrology Laboratory, RIKEN, Wako-shi, Saitama 351-0198,
Japan, \\
Innovative Space-Time Project, ERATO, Japan Science and Technology
Agency, Bunkyo-ku, Tokyo 113-8656, Japan and \\ 
Department of Applied Physics, Graduate School of
  Engineering, The University of Tokyo, Bunkyo-ku, Tokyo 113-8656,
  Japan}

\date{\today}

\begin{abstract}
We study the Ho$^{14+}$ ion as a candidate for extremely accurate and stable
optical atomic clock which is sensitive to the time variation of the
fine structure constant. We demonstrate that the proposed system has
all desired features including 
relatively strong optical electric dipole and magnetic dipole
transitions which can be used for cooling and detection. 
Zero quadrupole moments for the relevant states in the
clock  transition allows interrogating multiple ions to improve the
clock stability.
\end{abstract} 
\maketitle
%***************************************************************************

%\section{introduction}

 \begin{figure}
\epsfig{figure=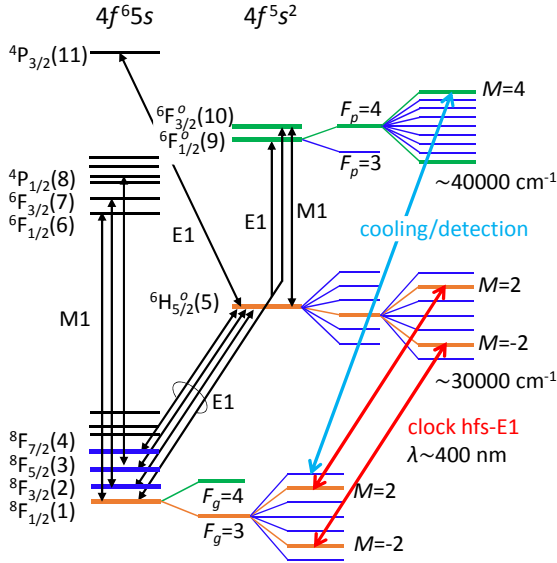,scale=0.65}
  \caption{Low-lying energy levels of Ho$^{14+}$. Numbers in parentheses correspond to those in Table II and IV to denote electronic states.}
%  \caption{\normalsize{Low-lying energy levels of Ho$^{14+}$.}
%  \caption{Low-lying energy levels of Ho$^{14+}$.}
    \label{fig1}
\end{figure}

Theories unifying gravity with other fundamental interactions suggest
a possibility for fundamental constants to vary in space or time (see,
e.g. \cite{Uzan}). Searching for variation of the fundamental constants 
is an important way of testing these theories and finding new physics
beyond standard model. Laboratory measurements limit the rate at which
the fine structure constant $\alpha$ ($\alpha = e^2/\hbar c$) varies
in time to about $10^{-17}$ per year~\cite{Al+,Dy}. On the other hand,
the analysis of quasar absorption spectra shows that the fine structure 
constant  may vary on astronomical scale along a certain direction in space
forming the so called alpha-dipole~\cite{alpha-dipole}. Earth
movement   in the framework of the alpha-dipole 
%???
(towards area of bigger alpha)
would lead to the local time
variation of $\alpha$ on the scale of $10^{-19}$ per year~\cite{BF}. To
test the alpha-dipole hypothesis in terrestrial studies one needs
a better  accuracy in the laboratory measurements.
%??? 
%This can be probably achieved with the use of optical atomic clocks. Indeed,
The current best limit on the local time variation of $\alpha$ has been
obtained by comparing Al$^+$ and Hg$^+$ optical
clocks~\cite{Al+}. The result of this comparison limits time
variation of alpha to  ($\partial
\alpha/\partial t)/\alpha = (-1.6 \pm 2.3) \times 10^{-17} \
\rm{yr}^{-1}$~\cite{Al+}), i.e. about two orders of magnitude
improvement in accuracy is needed to test the alpha-dipole
hypothesis. 

Current-best optical-clocks approach fractional uncertainty %HK 
of 
$\sim 10^{-18}$~\cite{Yb-clock,Sr-clock,Katori}. However, %the frequencies of 
the transitions used in these clocks are not sufficiently sensitive
to the variation of $\alpha$~\cite{two-el-at}. If $\alpha$ does change
in time, the relative change $\delta \omega/\omega$ of the clock frequencies in any given period of
time would be about an order of magnitude smaller than relative change in $ \delta \alpha/\alpha$.
%of $\alpha$.
% ($\delta \omega/\omega \sim 0.1 \delta \alpha/\alpha$).
%the change of frequency goes by about an order of magnitude
%slower than change of $\alpha$.

\begin{table}
\caption{Highly charged ions with optical clock transitions sensitive to
  variation of $\alpha$ satisfying the criteria discussed in the text.}
\label{t:ions}
\begin{ruledtabular}
\begin{tabular}{llc lcr}
\multicolumn{1}{c}{Ion} & 
\multicolumn{2}{c}{Ground state} & 
\multicolumn{2}{c}{Clock state} & 
\multicolumn{1}{c}{Energy (cm$^{-1}$)}\\
\hline
Er$^{14+}$ & $4f^75s$ & $J=4$ &  $4f^65s^2$ & $J=0$ & 18555 \\ 
Dy$^{10+}$ & $4f^75p$ & $J=3$ &  $4f^65p^2$ & $J=0$ & 20835 \\ 
%Dy$^{14+}$ & $4f^55s$ & $J=2$ & $4f^45s^2$ & $J=4$ & 33663 \\ 
%Dy$^{14+}$ & $4f^55s$ & $J=2$ & $4f^6$ & $J=0$ & 34737 \\ 
Ho$^{14+}$ & $4f^65s$ & $J=0.5$ & $4f^55s^2$ & $J=2.5$ & 23800 \\ 
\end{tabular}
\end{ruledtabular}
\end{table}

It has been suggested in \cite{hci1} to use highly-charged ions (HCI)
as extremely accurate atomic clocks to probe possible variation of
$\alpha$. 
While HCIs %HK: as general ions are not so
 are less  sensitive to external perturbations due to 
their compact size, their sensitivity to %HK
variation of $\alpha$ is enhanced due to larger relativistic effects.
In spite of the general trend of increasing energy intervals in HCIs, %HK 
it is still possible to find clock transitions which are in optical range. 
This is due to electron energy level crossing while
moving from the Modelung to the Coulomb level ordering with increasing of
the ionization degree~\cite{crossing}. A number of optical clock
transitions sensitive to variation of $\alpha$ were found and
examined in
Refs.~\cite{hci1,crossing,hole,simple1,simple2,simple3,simple4,simple5}. 

Proposals for HCI clocks so far focused on their high accuracy
aspects, such as the reduced sensitivity to blackbody radiation shift
due to lack of infrared and optical transitions~\cite{simple2}. These
proposals, therefore, naturally assume to use quantum logic
spectroscopy (QLS)~\cite{QLS} by co-trapping optically accessible
ions. From an experimental viewpoint, however, increase of number of ions
is a pressing need in order to improve clock stability, which is limited by the
quantum projection noise (QPN)~\cite{Itano}. Clock stability at QPN limit  %HK deleted fluctuation
is given by
$\sigma_y \approx \frac{1}{Q}\frac{1}{\sqrt{N\tau/\tau_c}}$,
where $Q = \omega/\Delta\omega$ is the effective quality-factor for the
interrogated transition and $N$ is the number of ions measured in a
clock cycle time $\tau_c$. Assuming $Q \approx 10^{15}$ and
$\tau_c=1$~s that are achievable with the-state-of-the-art
lasers~\cite{Kessler}, a goal uncertainty of e.g., $1\times 10^{-18}$
determines necessary averaging time of $\tau = 10^6N^{-1}$~s. In order 
to achieve this goal within a realistic experimental run time of $\tau
\sim 10^4$~s ($\sim 3$~hours), %HK as many as 
$N \approx 10^2$ ions need to be
interrogated simultaneously. Regrettably, QLS is not applicable to
such number of ions because of difficulties in controlling degrees of
motional states of the ions; therefore %HK, and 
electron shelving
detection~\cite{Nagourney} needs to be adopted, instead. 
Moreover, with $N>1$ ions present, %HK
spatially
inhomogeneous quadrupole-shift originating from coulomb interactions
between ions becomes a major concern.  

For these reasons we relax the condition of having simple electron
structure and search for ions which have relatively strong electric
dipole (E1) or magnetic dipole (M1) optical transitions from both
ground and clock states. The criteria for a suitable %HK 
system are formulated
as following
\begin{itemize}
\item Clock transition has high  sensitivity to the variation of
  $\alpha$ (e.g. it is a $5s$ to $4f$ or $5p$ to $4f$ transition).
\item It is optical transition ($230 \ \rm{nm} < \lambda < 2000 \ \rm{nm}$
  or $5000 \ \rm{cm}^{-1} < \hbar \omega < 43000 \ \rm{cm}^{-1}$) for existing narrow-linewidth lasers to access.
\item It is a transition between long-lived states %HK:from ground to an excited metastable state
  with lifetime between 100 and $10^4$ seconds.
\item There are other relatively strong optical transitions
  (equivalent lifetime $\tau \le 1$ ms).
\item Clock transition is not sensitive to the gradients of electric
  field  (e.g, there is no electric  quadrupole moment in both states).  
\end{itemize}
Note that blackbody radiation shift is not a limiting factor for HCI
clocks since trapping of HCI for more than an hour requires ultrahigh vacuum environment
that is %only 
achievable with cryogenic pumps, which naturally cool
environmental temperature to $ \sim$ 10~K.

Clock transitions for some HCIs that %HK:come very close to 
satisfy all these criteria are presented in Table~\ref{t:ions}. All these ions
have complicated electronic  structure with six to eight electrons in
open $4f$, $5s$ and $5p$ shells. 
The analysis is done by the many-electron version of the configuration
interaction (CI) method developed in Ref.~\cite{CI}. Breit and quantum
radiative corrections are also included as has been described in
Ref.~\cite{Breit,QED}. The precision of this analysis is limited, which %HK
means that %the values of 
the clock 
frequencies calculated in Table~\ref{t:ions} may differ from real ones by up
to estimated value of 10000 cm$^{-1}$. Even with this shift most of the
presented transitions remain in the optical range. 

\begin{table} %HK modified
\caption{Low lying states of Ho$^{14+}$, their energies (cm$^{-1}$),
  $g$-factors and lifetimes (s). %HK: Please check(ms). 
Not all states are shown but only
the   ground state, the clock state (at $E$=23823 cm$^{-1}$), and states which
  have relatively strong E1 or M1 transitions to either of the two states.
Electronic states are numerated as $n=1,2,3...$ from the lowest state. %HK 
} 
\label{t:EL}
\begin{ruledtabular}
\begin{tabular}{rll rdl}
\multicolumn{3}{c}{State} & 
\multicolumn{1}{c}{Energy} & 
\multicolumn{1}{c}{$g$-factor} & 
\multicolumn{1}{c}{lifetime} \\ 
$n$ & & & \multicolumn{1}{c}{(cm$^{-1}$)} & & %HK: N conflicts with atom number
\multicolumn{1}{c}{(s)} \\ %HK: not[] for units
\hline
1 & $4f^6 5s$ & $^8$F$_{1/2}$ &        0 &  3.94 & \\
2 &           & $^8$F$_{3/2}$ &     1013 &  1.99 & 8.2 \\
3 &           & $^8$F$_{5/2}$ &     2499 &  1.71 & 2.3 \\
4 &           & $^8$F$_{7/2}$ &     4298 &  1.61 & 1.4 \\

5 & $4f^5 5s^2$ & $^6$H$^o_{5/2}$ & 23823 &  0.286 & 37 \\

6 & $4f^6 5s$ & $^6$F$_{1/2}$ &    30199   & -0.17 & 0.10 \\
7 &           & $^6$F$_{3/2}$ &    31713   & 1.22 & 0.14 \\
8 &           & $^4$P$_{1/2}$ &    34285   &  2.96 & 0.01 \\
%          & $^4$P$_{3/2}$ &    38024   &  1.80 & 0.05 \\
                                
9 & $4f^5 5s^2$ & $^6$F$^o_{1/2}$ & 37351 &  -0.63 & 0.002  \\
10 & $4f^5 5s^2$ & $^6$F$^o_{3/2}$ & 37771 &  1.05 & 0.005  \\

11 & $4f^6 5s$ & $^4$P$_{3/2}$ &    38023   & 1.80 & 0.01 \\

%          & $^6$G$_{3/2}$ &    50850   & 0.04 &  \\
%          & $^6$G$_{5/2}$ &    51043   & 0.80 &  \\
%          & $^6$G$_{7/2}$ &    51279   & 1.08 &  \\

\end{tabular}
\end{ruledtabular}
\end{table}

We now discuss in more detail the Ho$^{14+}$ ion which currently
appears as the best candidate for an optical clock. Its level structure is
shown in Fig.~\ref{fig1}. Table~\ref{t:EL} shows low-lying energy
levels of Ho$^{14+}$. For the $4f^65s \ ^8$F$_{1/2}$ ground
state, the $4f^5 5s^2 \ ^6$H$^o_{5/2}$ state can be taken as the clock state. %HK modified
%Table~\ref{t:EL} shows energy levels of only those states 
Only levels which are connected to the ground or clock state by
electric dipole (E1) or magnetic dipole (M1) transitions are shown in
 Table II. The total number of states in the  
considered energy interval is much larger. 
For each fine structure 
multiplet, $N_s = \min(2S+1,2L+1)$ gives the number of states, %$N_s$ is given by 
where $S$ is total spin and $L$ is total
orbital angular momentum. The values of $S$ and $L$ and corresponding 
names of the multiplets are found from the values of the $g$-factors 
(see Table~\ref{t:EL}) 
\begin{equation}
g = 1 + \frac{J(J+1) - L(L+1) + S(S+1)}{2J(J+1)}.
\label{eq:g}
\end{equation}
For example, the ground state is the lowest state 
of the $^8$F multiplet which has seven levels with total angular momentum 
$J$ ranging from $J=1/2$ to $J=13/2$. Corresponding non-relativistic 
value of the $g$-factor (\ref{eq:g}) is $g=4.0$ while the value
obtained in the CI calculations is $g=3.94$.

The clock transition can go as a magnetic quadrupole (M2) or electric
octupole (E3) transition. Corresponding rates are $8\times 10^{-14}$
s$^{-1}$ and $4\times 10^{-12}$ s$^{-1}$.
However, transition
rate is dominated by the electric dipole (E1) transition mediated by the
magnetic-dipole hyperfine-structure (hfs) interaction. Holmium atom
has single stable isotope $^{165}$Ho with nuclear spin $I=7/2$,
the nuclear magnetic moment $\mu = 4.173\mu_N$~\cite{web} and the nuclear
electric quadrupole moment $Q=3.58$b~\cite{Pekka}
(1b=$10^{-28}$m$^2$). Calculated hfs 
constants $A$ and $B$ are presented in Table~\ref{t:hfs}.

\begin{table}
\caption{Magnetic dipole and electric quadrupole hfs constants $A$ and
  $B$ for the ground and clock states of $^{165}$Ho$^{14+}$.} 
\label{t:hfs}
\begin{ruledtabular}
\begin{tabular}{ll cc}
\multicolumn{2}{c}{State} & 
\multicolumn{1}{c}{$A$} & 
\multicolumn{1}{c}{$B$} \\ 
 & & \multicolumn{1}{c}{(GHz)} & 
\multicolumn{1}{c}{(GHz)} \\ 
\hline
$4f^6 5s$   & $^8$F$_{1/2}$  &  96.5 & 0 \\ 
$4f^5 5s^2$ & $^6$H$^o_{5/2}$ & 3.53 & -6.04 \\
\end{tabular}
\end{ruledtabular}
\end{table}

The largest contribution to the hfs-mediated E1 clock transition
moment %HK: Is this ok?amplitude 
is given by
\begin{eqnarray}
d_{\rm hfs-E1} = \frac{\langle 4f^6 5s \ ^8{\rm F}_{1/2}|\hat {\rm H}_{\rm hfs}|4f^6 5s \ %HK: operator may be in roman
  ^8{\rm F}_{3/2} \rangle}{\Delta E} \\%HK: If this is transition matrix element \mu is more visible
\times \langle 4f^6 5s \ ^8{\rm F}_{3/2}|E1|4f^5 5s^2
  \ ^6{\rm H}^o_{5/2} \rangle, \nonumber
\label{A-E1-hfs}
\end{eqnarray}
where $\hat {\rm H}_{\rm hfs}$ is magnetic-dipole hfs Hamiltonian, $E1$ is the
electric-dipole operator, and $\Delta E = 1013$ cm$^{-1}$ is the
energy interval between ground and dominating intermediate state. This
term dominates due to small energy denominator. The hfs matrix element
is equal to $1.94 \times 10^{-5}$ a.u. The E1 matrix element is
related to the electric dipole transition amplitude between  $n=2$ and 5 states %HK
 (see Table~\ref{t:EL}) by an angular coefficient. Its value
$(~10^{-5}$ a.u.) is small due to absence of the single-electron
electric dipole transition between leading configurations (the $5s$ -
$4f$ transition cannot go as E1). The amplitude is not zero due to
configuration mixing. 

Spontaneous emission rate corresponding to the hfs-E1 transition-moment %HK
[see Eq.~(2)] is the same as that for the usual E1 transition (we use atomic units) %HK: this eq should be (2) but \ref{A-E1-hfs} cite (3) I do not know the reason.
\begin{equation}
A_{\rm hfs-E1} = \frac{4}{3}(\omega\alpha)^3\frac{d_{\rm
    hfs-E1}^2}{2F_c+1},%HK: A reminds me of Einstein A coeff.
\label{T-E1-hfs}
\end{equation}
where $\omega = 0.092$ a.u. is the frequency of the clock transition, 
$d_{\rm hfs-E1}$ is reduced matrix element,
and $F_c$ is the total angular momentum of the clock state
($\mathbf{F} = \mathbf{I} + \mathbf{J}$). 
%The transition rate between
%states of $F_g=3$ in the ground state and $F_c=3$ in the clock state
The emission rate from the clock state with $F_c=3$ is
$7.5 \times 10^{-7}$ s$^{-1}$.
% which corresponds to the partial lifetime of $1.3 \times 10^{6}$ s. 
Note that this transition rate %HK added
 gives negligible contribution to the clock state
lifetime, which is dominated by the E1 %HK
transitions to three first excited states of the ground state
configuration. %HK (see below) cites too far.  

Rabi frequency $\Omega=d_{\rm hfs-E1}\cal{E}$ %HK: this would be more visible for readers
for the $4f^6 5s \ {}^8{\rm F}_{1/2}\rightarrow4f^5 5s^2\, {}^6{\rm H}^o_{5/2}$ clock transition %HK
%between clock (5) and ground (1) states 
is estimated from  
Eq.~(2) and laser electric field $\cal{E}$. %HK this should be (2) \ref{A-E1-hfs}
%We find that it takes 1.6 W/m$^2$ 
For laser intensity of 16 mW/cm$^2$, the Rabi frequency is estimated to be 
1 Hz. %of Rabi oscillations 1Hz corresponds to Q assumption before. 
Corresponding frequency shift of the
clock transition is about $10^{-3}$ Hz. %HK this is correspondingly modified. please check.
%At laser intensity of $2\times 10^6$ W/m$^2$ (20 mW/(0.1 mm)$^2$) the shift of clock frequency is 12 Hz and
%frequency of Rabi oscillations is 111 Hz. 

The ${}^6{\rm H}^o_{5/2}$ clock state is split by the hfs interaction into six levels with
$F_c=1, \dots , 6$. 
Similarly, the ${}^8{\rm F}_{1/2}$  ground state is split to $F_g=3, 4$.
We consider $F_c=F_g=3$ with $M=2$ states as a clock %HK: added
transition, because it is insensitive to the  electric-field gradients that couple 
to atomic quadrupole moment and cause quadrupole shift. %HK reference may be added
Since the quadrupole moment is proportional to $3M^2 - F(F+1)$
($M$ is projection of $F$), choosing $F=3$ and $M= \pm 2$ eliminates
quadrupole frequency shift. 
Note that the ground state has no
quadrupole moment because of the total electron angular %HK
momentum $J=1/2$.  

Optical pumping %HK:Preparation of ions in 
to the desired $M$ states can be performed %by optical pumping %the electronic states of ions 
by employing strong E1 transitions from
the ground state as  given in Table~\ref{t:tr}. 
%by applying  in a bias magnetic field. By
In a presence of a bias magnetic field, %by applying 
applying $\sigma^+$ polarized light resonant to 
%the $F_g=3$ in the ground state (1) to  
the $4f^55s^2\ {}^6{\rm F}^o_{1/2}(F_p=4)$ pump state (denoted by $n=9$), ions are
spin-polarized in the $F_g=M=3$ state, which is then transfered
to the $F_g=3$, $M=2$ state by a stimulated Raman adiabatic
passage via the pump state.  
The same E1 line can be used for the electron-shelving  detection~\cite{Nagourney}. % can be performed on. %in a reversed way: 
With $\sigma^+$ polarized light resonant to this pump transition, %After transferring $F_g=3$, $M=2$ state to $F_g=M=3$ state, 
the ground state population is swept into the $F_g=M=3 \rightarrow F_p=M=4$ quasi-cyclic transition, where four recycling lasers may depopulate the $F_g=4$ state as well as $F=3,4,5$ states in the ${}^8{\rm F}_{3/2}$ manifold. % will be applied.  
The same scheme as above can be conveniently used to  sideband-cool a chain of ions. 
Here we assume Ho$^{14+}$ ions are initially cooperatively-cooled by coolant ions such as Be$^+$ whose mass-charge  ratio ($A/q=9$) is similar to that of  Ho$^{14+}$ ($A/q\approx 11.8$).  

\begin{table}
\caption{Electric dipole (E1) and magnetic dipole (M1) transitions
  involving ground and clock states including %HK cooling state is named as pumping state
  spontaneous emission  rates from upper state ($n_U$) to lower state ($n_L$), where numeration of the states
  corresponds to Table~\ref{t:EL}. Numbers in square brackets
  represent powers of ten.}
% {\bf [Comment: HK thinks Energy column is not necessary and better to be deleted, as $N_U$ is inserted now and Energy and Frequency in the same units are confusing as both of them are energy.]}} 
\label{t:tr}
\begin{ruledtabular}
\begin{tabular}{ccrcc}
\multicolumn{2}{c}{Transition} & 
%\multicolumn{1}{c}{Energy} & 
\multicolumn{1}{c}{Frequency} &
\multicolumn{1}{c}{Amplitude} &
\multicolumn{1}{c}{Rate} \\
$n_U \rightarrow n_L$ & \multicolumn{1}{c}{Type} & 
%\multicolumn{1}{c}{(cm$^{-1}$)} &
\multicolumn{1}{c}{(cm$^{-1}$)} &
\multicolumn{1}{c}{($10^{-3}$ a.u.)} &
\multicolumn{1}{c}{(s$^{-1}$)} \\ 
\hline
                  &          &       &      &     \\
\multicolumn{5}{c}{a. Transitions to ground state ($n_L=1$)} \\ 
                  &          &       &      &     \\
$2 \rightarrow 1$ & M1  &  1013 & 15 &  0.12\\
$6 \rightarrow 1$ & M1  & 30199 & 0.48 & 6.4 \\
$7 \rightarrow 1$ & M1  & 31713 & 0.29 & 1.4 \\
$8 \rightarrow 1$ & M1  & 34285 & 0.67 & 18 \\
$9 \rightarrow 1$ & E1  & 37351 & 2.81 & 410 \\
$10 \rightarrow 1$ & E1  & 37771 & 2.18 & 130 \\
$11 \rightarrow 1$ & M1 & 38023 & 0.49 & 6.7 \\
                  &    &              &      &     \\
\multicolumn{5}{c}{b. Transitions from/to clock state ($n_{U/L}=5$)} \\  
                  &      &              &      &     \\
$5 \rightarrow 2$ & E1  & 22810 & 5.1[-2] & 1.0[-2] \\
$5 \rightarrow 3$ & E1  & 21324 & 4.3[-3] & 6.0[-5] \\
$5 \rightarrow 4$ & E1  & 19525 & 8.1[-2] & 1.6[-2] \\
$10 \rightarrow 5$ & M1  & 13948 & 4.6[-1] & 0.29 \\
$11 \rightarrow 5$ & E1  & 14200 & 7.1[-2] & 7.0[-3] \\
%                  &    &       &       &      &     \\
%\multicolumn{6}{c}{c. Transitions involving cooling state ($N_U=6$%$E$=30199 cm$^{-1}$
%)} \\ 
%                  &    &       &       &      &     \\
%$6 \rightarrow 1$ & M1 &     0 & 30199 & 0.48 & 6.4 \\
%$6 \rightarrow 2$ & M1 &  1013 & 29186 & 0.37  & 3.4 \\
\end{tabular}
\end{ruledtabular}
\end{table}

Lifetime of the clock state is about 37~s. It is mainly due to
electric dipole transitions to the first three excited states in the
ground state configurations (see Table~\ref{t:tr}). 
%Relative linewidth $\Gamma/(\hbar\omega) = 5 \times 10^{-18}$.
%The strongest E1 and M1 transitions from
%ground and clock states and the state with $E=30199$~cm$^{-1}$ which
%can be used as cooling state are
The table also lists E1 and M1 transitions that can be used for electronic state manipulation as well as laser-cooling  ions' motion. 
In particular, the ${}^6{\rm H}^o_{5/2}\rightarrow {}^6{\rm F}^o_{3/2}$ (M1) transition followed by an E1 decay to the ground state may be conveniently used to depopulate the clock state.
The amplitudes are given by reduced matrix elements of the electric dipole
($e\mathbf{r}\cdot\mathbf{E}$) and magnetic dipole
($\mathbf{\mu}\cdot\mathbf{H}$) operators.
The M1 amplitudes include electron magnetic moment $\mu$ which in
atomic units is equal to $\alpha/2 = 3.65 \times 10^{-3}$
(Gaussian-based atomic units). 
Spontaneous emission rates for  
both E1 and M1 transitions are given by Eq.~(\ref{T-E1-hfs}) with $F_c$
replaced by $J$.
%The transitions which determine the lifetime of the clock state are given
%in first three lines of the Table~\ref{t:tr}b. 
%Othertransitions might be used for detection, cooling and populating of the clock state. 
%The transitions of particular interest are the relatively
%strong E1-transition from the ground state to the $^6$F$^o_{1/2}$
%state of energy $E$ = 37351 cm$^{-1}$ and M1-transition from the
%ground state to the $^6$F$_{1/2}$ state of energy $E$ =
%30199~cm$^{-1}$. Either of these states can be used for cooling. 
%In either case the $^8$F$_{1/2}$, $F=3$ to $^6$F$_{1/2}$, $F=4$
%transition is to be used as a quasi-cyclic one. The repumping from the
%$^8$F$_{1/2}$, $F=4$ and $^8$F$_{3/2}$, $F=3,4,5$ is needed which
%would require four recycling lasers.

The clock transition is sensitive to variation of the fine structure
constant since it is a $4f$ to $5s$ transition~\cite{crossing}. It is
convenient to present frequency dependence on $\alpha$ in the form
\begin{equation}
\omega(x) = \omega_0 + qx, \ {\rm where} \ x
= \left[\left(\frac{\alpha}{\alpha_0}\right)^2 -1 \right].
\label{eq:wq}
\end{equation}
The sensitivity coefficient $q$ is found in atomic calculations by
varying the value of $\alpha$ in computer codes.
%\begin{equation}
%q = \frac{\omega(0.01) - \omega(-0.01)}{0.02}.
%\label{eq:q}
%\end{equation}
In our calculations $q$ = -186000 cm$^{-1}$ and variation of the clock
frequency is related to the variation of the fine structure constant
by
\begin{equation}
\frac{\delta\omega}{\omega} =-18 \frac{\delta\alpha}{\alpha}.
\label{eq:alpha}
\end{equation}
As it has been mentioned above the precision of present calculations
based on the version of the CI method developed in Ref.~\cite{CI} is
limited. The main problem is the energy interval between states of
different configurations. Precision for energy intervals within one
configuration is better and is on the level of 10 to 20\%. Therefore, to start with, 
it is  important to spectroscopically investigate %spectrum start  spectroscopic measurements for %from measuring frequency of 
transitions between different configurations. The best candidates % in thecase of Ho$^{14+}$ ion 
are the E1 transitions from the ground state to the  ${}^6F^o_{1/2,3/2}$
states located at  $\sim 4\times 10^4$ cm$^{-1}$.
Once this
frequency is measured, it  helps to adjust the calculations and
improves significantly all other predictions. 
% $E \approx 37351$ cm$^{-1}$ and $E \approx 37771$
Note that even significant 
change in the frequency of the clock transition of Ho$^{14+}$ leaves
it in optical range making the ion an attractive candidate for
experimental study.

%\begin{acknowledgments}

The work was supported by the Australian Research Council.

%\end{acknowledgments}

\end{document}